\documentclass[]{spie}  

\usepackage{subfig}
\usepackage{amsmath,amsfonts,amssymb}
\usepackage{graphicx}
\usepackage{multicol}
\usepackage[colorlinks=true, allcolors=blue]{hyperref}
\usepackage{mwe}
\usepackage{tabularx}
\usepackage[inline]{enumitem}
\newlength{\characterlength}
\usepackage{xcolor}

\newcommand{\editsA}[1] {\textcolor{black}{#1}}
\title{Predictive wavefront control on Keck II adaptive optics bench: on-sky coronagraphic results}

\author[a]{Maaike A.M. van Kooten}
\author[a]{Rebecca Jensen-Clem}
\author[b,c]{Sylvain Cetre}
\author[b]{Sam Ragland}
\author[d]{Charlotte Z. Bond}
\author[a]{J. Fowler}
\author[b]{Peter Wizinowich}
\affil[a]{University of California Santa Cruz, 156 High Street
Santa Cruz, CA, USA}
\affil[b]{W. M. Keck Observatory, 65-1120 Mamalahoa Hwy, Waimea, HI, USA}
\affil[c]{University of California 
Observatories,
1156 High Street
Santa Cruz, CA, USA }
\affil[d]{The UK Astronomy Technology Centre, Royal Observatory Edinburgh, Edinburgh, UK}

\authorinfo{Further author information: (Send correspondence to Maaike A.M van Kooten)\\ E-mail: mvankoot@ucsc.edu  }

\begin{document} 
\maketitle

\begin{abstract}
The behavior of an adaptive optics (AO) system for ground-based high contrast imaging (HCI) dictates the achievable contrast of the instrument. In conditions where the coherence time of the atmosphere is short compared to the speed of the AO system, the servo-lag error can become the dominant error term of the AO system. While the AO system measures the wavefront error and subsequently applies a correction (typically taking a total of one or a few milliseconds), the atmospheric turbulence above the telescope has changed resulting in the servo-lag error. In addition to reducing the Strehl ratio, the servo-lag error causes a build-up of speckles along the direction of the dominant wind vector in the coronagraphic image, severely limiting the contrast at small angular separations. One strategy to mitigate this problem is to predict the evolution of the turbulence over the delay time. Our predictive wavefront control algorithm minimizes, in a mean square sense, the wavefront error over the delay and has been implemented on the Keck II AO bench. In this paper, we report on the latest results of our algorithm and discuss updates to the algorithm itself. We explore how to tune various filter parameters based on both daytime laboratory tests and on-sky tests. We show a reduction in residual-mean-square wavefront error for the predictor compared to the leaky integrator (the standard controller for Keck) implemented on Keck for three separate nights. Finally, we present contrast improvements for daytime and on-sky tests for the first time. Using the L-band vortex coronagraph for Keck’s NIRC2 instrument, we find a contrast gain of up to 2 at a separation of 3 $\lambda/D$ and up to 3 for larger separations (3-7$\lambda/D$). 

\end{abstract}


\section{INTRODUCTION}
\label{sec:intro}  

Single conjugate adaptive optics (AO) systems provide a good correction over a modest field-of-view, thereby enabling a vast array of science. Within a high contrast imaging (HCI) system, this simplest form of AO becomes a powerful tool when observing bright, on-axis guide stars. By delivering an extreme AO correction through high speed operation and maximizing the actuator density of the deformable mirror (DM), HCI systems can search for exoplanets and disks around bright stars. The resulting AO systems provide Strehl ratios (SR) of greater than 60\% in near-infrared (NIR) wavelengths~\cite{Bond_2020, Macintosh_2014}. 

With these advancements in AO, HCI systems have reached post-processed contrasts of $10^{-6}$ at spatial separations of 200 milli-arc-seconds~\cite{zurlo_2016}. Closer to the host star, a typical AO system is dominated by either non-common path aberrations or the servo-lag error. The servo-lag error, in the worst-case scenario, can visually be seen as a wind-driven halo in a coronagraphic image for instruments such as the Very Large Telescope's (VLT) SPHERE HCI instrument, contaminating the region near the inner working angle of the coronagraph~\cite{Cantalloube_2018}. For other HCI systems, the servo-lag error is a more subtle build-up of speckles. The servo-lag error is due to the atmospheric turbulence above the telescope evolving in the time between when the wavefront is measured and the new DM commands are applied. The delay time includes \editsA{half} the wavefront sensor (WFS) integration time, the wavefront sensor camera read-out time, computation time, and the DM response time; it ranges from 1-4 wavefront sensor frames, depending on the system. One solution to minimize the servo-lag error is to predict the evolution of the wavefront over the known delay using predictive control. 

Over the last five years, renewed efforts to implement predictive control on-sky for HCI have resulted in algorithms such as empirical orthogonal functions (EOF)~\cite{Guyon_2017} which is being tested at SCExAO~\cite{Nem_2015,cacao} and W. M. Keck Observatory ~\cite{Jensen_2019}. Van Kooten et al. 2017, 2019~\cite{Maaike_2017,vanKooten_2019} proposed a recursive solution to the minimum mean square cost function and within their framework test prediction for non-stationary turbulence as well as VLT/SPHERE telemetry~\cite{vanKooten_2020}. Most recently, data-driven subspace predictive control~\cite{Haffert_2021} has been proposed to operate in closed-loop, with plans to test on-sky with the MagAOX system~\cite{MagAOX}. \editsA{This work differs from previous work in the early 2000s focused on optimal control algorithms such as Linear-Quadratic-Gaussian control (LQG)~\cite{Roux_2004,Petit_2008,Petit_2009}, H2 optimal control~\cite{Hinnen_07}, and Predictive Fourier control~\cite{Poyneer_2007} that all rely on a combination of modeling and data-driven methods to identify input disturbances (both from telescope vibrations and turbulence) to predict the disturbances. Some of these methods have been successfully commissioned for tip-tilt control including on SPHERE/Very Large Telescope~\cite{Beuzit_2019}, and GPI/Gemini Telescope~\cite{Poyneer_2016} and are being tested for other low order modes on-sky~\cite{Sinquin_2020}. Finally, efforts to implement machine learning for predictive control are underway various groups. Early results from CANARY showed machine learning algorithms on-sky as a reconstructor~\cite{Osborn_2014_onsky_canary}. More recent simulation work show promising results when comparing EOF to a neural network under certain assumptions and advocate for operating non-linear wavefront sensors~\cite{Wong_2021}. However, more work into necessary to understand the require training for on-sky deployment and the required hardware for real-time-control. These are also discussed by J. Nousianen et al (2021)~\cite{Nousiainen_2021} and R. Swanson et al. (2021)~\cite{Swanson_2021} while showing results for other machine learning algorithms.}

In this paper, we focus on W. M. Keck Observatory's HCI system, Keck II's near-infrared camera (NIRC2). In Xuan et al. 2018~\cite{2018AJ....156..156X}, the authors found a correlation between the final post-processed contrast for NIRC2 and the coherence time of the atmosphere divided by the WFS integration time when using the Shack-Hartmann wavefront sensor (SHWFS); both variables influence the impact of the servo-lag error. These results show that predictive control would improve the performance of the system. Predictive control was implemented on the Keck II AO bench and Jensen-Clem et al. 2019~\cite{Jensen_2019} show a factor of 2.2 reduction in the residual-mean-square (RMS) error during daytime tests with simulated turbulence projected onto the DM. Since then, updates to the AO system have been made along with the predictive controller. In this paper, we present the latest results from daytime tests and on-sky engineering time at Keck, including a new recursive implementation tested during the day and contrast measurements from both day and night time tests using EOF. \editsA{We evaluate the performance of the new controller by looking at the residual wavefront error but also the contrast curve for NIRC2 specifically looking for improved contrast at small angular separations where the servo-lag error is expected to have  the most impact.}

We outline the paper as follows: an overview of Keck II AO and the NIRC2 instrument is given in Sec.~\ref{sec:Keck_AO_overview}. In Sec.~\ref{sec:controller}, we give details of the controller implemented on Keck including the predictive control (Sec.~\ref{sec:pred_control}). We present results from daytime and night time tests in Secs.~\ref{sec:daytime} and~\ref{sec:nighttime}, respectively. Discussion of the results are given in Sec.~\ref{sec:discussion}. We outline our plans for the system in Sec.~\ref{sec:future}. Finally, we end with our summary and conclusions in Sec.~\ref{sec:conclusions}. 

\section{Keck II Adaptive Optics system and NIRC2}
\label{sec:Keck_AO_overview}  
NIRC2 (PI: Keith Matthews) contains two vortex coronagraphs to cover the full wavelength range of the instrument (K-, and L'-/M-band vortex~\cite{L_band_vortex}, respectively) to enable HCI. NIRC2 is fed by the Keck II AO bench that operates with either a visible SHWFS~\cite{Wizinowich_2000} or a near-infrared pyramid wavefront sensor (PyWFS)~\cite{Bond_2018,Bond_2020}. The PyWFS enables AO correction for much fainter, redder guide stars, supporting unique science capabilities. For the remainder of this paper we will be referring to the PyWFS configured AO system. 

The PyWFS controls the 21x21 actuator (total of 349 actuators) Xinetics DM having 40 pixels across the pupil, and operating with a modulation radius of 5 $\frac{\lambda}{D}$. The real-time computer (RTC) for the PyWFS makes use of the CACAO framework~\cite{cacao,Cetre_2018} allowing the system to operate at 1kHz. Simulations and on-sky commissioning have verified that the PyWFS is able to achieve upwards of 65~\%  SR for an H-band magnitude of 5 and can provide AO correction up to H-band magnitude 12~\cite{Bond_2020}. The AO delay is determined to be 1.7 milli-seconds~\cite{Cetre_2018} when running the PyWFS at its full frame-rate. 
\section{Adaptive Optics Control}
\label{sec:controller}
\subsection{Integral control law}
Operating at 1kHz with 21 actuator across the aperture in H-band allows for the PyWFS to achieve good performance with the standard AO control law: a leaky integrator. In Eq.~\ref{eq:leaky_int}, the leaky integrator calculates the DM commands, $V^{int}(t)$, at time step $t$. 
\editsA{\begin{equation}
    V^{int}(t)=k V^{int}(t-1)-g(CM \times [S(t)-S_{ref}])
    \label{eq:leaky_int}
\end{equation}}
The leak, $k$, is nominally set to 0.99 for standard operation on Keck and for our tests. The gain ($g$) has a typical value of 0.4 which is adjusted depending on seeing conditions. $S(t)$ and $S_{ref}$ are the slopes measured by the PyWFS at time $t$ and the reference slopes encoding the non-common path aberrations for NIRC2, respectively. ~\editsA{Note we match the equations of Bond et al. (2020)~\cite{Bond_2020} and mimics the RTC implementation by matching the available telemetry data streams as also done by Poyneer et al. (2007)~\cite{Poyneer_2007}.} Finally, the command matrix, $CM$, not only converts the slopes to DM actuator voltage but also encodes the number of modes being used and any high- and low- order (LO and HO, respectively) modal gains. The modal gains are applied in Fourier space with the cut-off-frequency (CoF) indicating where the trade off between LO and HO modes occurs in units of $\lambda/D$. Typically, it is taken to be the modulation radius of the PyWFS. Unless specifically stated the HO and LO \editsA{modal gains} were set to 0.4 and 1 respectively and the CoF is equal to 5 which is the default modulation radius of the PyWFS. The PyWFS residuals wavefronts (in DM command space) are then defined to be $CM \times [S(t)-S_{ref}]$.  Finally, the total number of controlled Zernike modes can be varied as well, with 350 modes being the maximum correctable number of modes, once again this is all done within the $CM$. \editsA{Therefore within the $CM$, the modal gains are applied in Fourier space and then the total number of controllable modes is defined in Zernike modes before calculating the voltage map for the DM. } Within the RTC the full command matrix is also stored $CM_{full}$ \editsA{which contains all the Zernike modes used for calibration not just the controlled modes}. This is used to generate the $CM$ used by the controller that applies the correct total number of modes. For more details on the controller and the slopes computation refer to Bond et. al. 2020~\cite{Bond_2020}.

\subsection{Predictive Control}
\label{sec:pred_control}
Using the separation principle, our predictive controller is implemented in two steps: \begin{enumerate*} 
  
    \item predict the input disturbance (i.e. wavefront error due to atmospheric turbulence above the telescope) and
    \item control the system plant (i.e., DM).
\end{enumerate*}   
Within this flexible framework, we can implement different predictive algorithms. We are also able to pick various control laws including open-loop control or the already existing leaky integrator implemented for the PyWFS. 

In this work, we predict the DM commands one lag time into the future by finding a filter that uses a subset of previous measurements subject to some cost function. \editsA{We focus on data-driven approaches where we use recent measurements to determine the future state of the input disturbance. With that we rely on spatial and temporal relationships of the data which not only allow for us to predict but also help reduce other error terms in the system that is present in the data.} Since we aim to predict the full input disturbance (e.g., the wavefront due to turbulence), we must first reconstruct the pseudo-open loop (POL) wavefront as the PyWFS is downstream from the DM measuring the residual wavefront error. 

\subsubsection{Pseudo-open loop reconstruction}

Making use of the shared memory structure within the CACAO RTC framework, the POL DM commands are calculated using the current PyWFS residual wavefronts and the previous DM commands from lag frames behind. Only the modes controlled by the leaky integrator are reconstructed to form the PyWFS residuals in this step and therefore are reconstructed using the $CM$, the filtered command matrix. At any point, up to 120000 frames (2 minutes at 1 kHz) of filtered POL DM commands are available in the shared memory and can be saved on command. The POL DM commands can be converted to optical-path-difference (OPD) using $0.6\frac{\mu m}{V}$. For calculating the predictive filter we make use of these filtered POL DM commands.
 
\subsubsection{Predictive filter}

There are a few different flavors of predictive control implemented on the PyWFS RTC that all aim to minimize the same cost function: the linear minimum mean square error (LMMSE). We write the LMMSE cost function in Eq.~\ref{eq:cost_function} where a unique predictive filter, $F^i$ is determined for each mode or in our case DM actuator. 
\begin{equation}
    min_{F^i} < || F^ih(t) -w_i(t+\delta t) || ^2 >
    \label{eq:cost_function}
\end{equation}
\editsA{The filtered POL DM commands we want to predict (i.e., true wavefronts) are represented by $w_i$ at $\delta t$ steps into the future while $\widetilde{w}$ is our POL measured data with $dt$ being the time step between wavefront measurements. Our regressors are contained in the history vector $h(t)$, which is formed from the POL DM commands (Eq.~\ref{eq:history}) containing temporal order, $n$, of previous measurements (i.e.,  $\widetilde{w}$). Each measurement vector is of length $m$. } 

\editsA{\begin{equation}
    h(t) =
    \begin{bmatrix}
\widetilde{w}_0(t) \\
\dots \\
\widetilde{w}_{m-1}(t) \\
\widetilde{w}_0(t-dt) \\
\dots \\
\widetilde{w}_{m-1}(t-dt)  \\
\dots \\
\widetilde{w}_{m-1}(t-(n-1)dt) 
\end{bmatrix}
    \label{eq:history}
\end{equation}}

In 2019, EOF was implemented on the PyWFS RTC in python such that the predictive filter was calculated in a batch sense by storing up to $l$ wavefront sensor measurements~\cite{Jensen_2019}. In this implementation, the cost function is solved in a regularized least-squares sense as shown in Eq.~\ref{eq:EOF} where $\alpha$ is the regularization parameter and $I$ is the identity matrix. $P$ and $D$ contain $l$ pairs of the true wavefront ($w_i$) and history vectors, $h$, respectively. 

\begin{equation}
    F^i= PD^T(DD^T+\alpha I)^{-1}
    \label{eq:EOF}
\end{equation}

The matrix multiplication of the predictive filter has now been implemented in the GPU on the PyWFS RTC, allowing us to predict at the speed of the system, an upgrade since 2019~\cite{Jensen_2019}, and therefore run on-sky. The EOF filter, using Eq.~\ref{eq:EOF} is first calculated offline in python and then passed to the shared memory once prediction is enabled. At the moment, the filter takes 9 seconds to calculate using 1 minute of telemetry ($l$=60000 frames). We have also updated the filter such that the number of temporal orders (how many previous measurements used to form $h$), $n$, can vary. Previously, we had been limited to a temporal order of 10 but now have the option to vary this parameter. Currently, however, doing so greatly increase the computation time to minute timescales.  

More recently, the recursive LMMSE (RLMMSE) solution\cite{vanKooten_2019} has also been implemented on the Keck RTC within python. The RLMMSE allows for the system to continuously update the predictive filter, effectively tracking any changes in the input disturbance (atmospheric turbulence). The solution to the cost function, Eq.~\ref{eq:cost_function}, can be written in terms of the auto-covariance of $h$ and the cross-covariance between the $w_i$ and $h$ ($C_{hh}$ and $c_{hw_i}$, respectively) as shown in Eq.~\ref{eq:batch_LMMSE}. Using this equation, we can calculate the batch solution by once again storing up data and estimating the auto-covariance and the cross-covariance matrices. We can also build-up our covariances using Eqs.~\ref{eq:auto_update} and~\ref{eq:cross_update}. In practice, the filter first runs in the background, once it has initially converged using a few thousand frames (e.g., $l$=O(1000)), the filter is then sent to the shared memory and prediction is enabled. The filter can then update recursively as fast as possible within the python implementation (each calculation takes roughly 0.02 seconds giving us an update rate of 50 Hz) at the moment. 

\begin{equation}
F^i=C_{hh}^+c_{hw_i}    
\label{eq:batch_LMMSE}
\end{equation}

\begin{equation}
    \label{eq:auto_update}
    C_{hh}^+(t)=C_{hh}^+(t-1)-\frac{C_{hh}^+(t-1)h^T(t)C_{hh}^+(t-1)}{1+h^T(t)C_{hh}^+(t-1)h(t)}
\end{equation}

\begin{equation}
    \label{eq:cross_update}
    c_{hw_i}(t)=c_{hw_i}(t-1)+w_i(t)h^T(t)
\end{equation}

\subsubsection{Control Law}
Our current implementation of the predictive filter is a semi-open loop controller. The predicted wavefronts return the DM commands for correcting the full atmospheric turbulence, \editsA{$V^{pred}(t)$ $[= F^ih(t)]$}. The predicted commands can be mixed with the integrator to generate the signal sent to the DM, $V(t)$. This configuration is shown in Eq.~\ref{eq:open_loop}. The integral controller (the second term) is the same as Eq.~\ref{eq:leaky_int}. Now the slopes encode the residuals of the predictor.  We tune the mixing parameter $m$ and the gain $g$ to have stable correction where $m$ values between 0.1 and 0.75 provide stability with values closer to 0.5 providing the best correction. \editsA{In the end, this mimics a leaky integrator operating on the predicted values and residuals that allows for us to control the system plant, combining with our predictive filter to become our predictive controller.}
\begin{equation}
    V(t)=mV^{pred}(t)+(1-m)((k)V^{int}({t-1})-g(CM \times [S(t)-S_{ref}])
    \label{eq:open_loop}
\end{equation}

\subsection{Residual wavefront error}
To determine the performance of the AO system we look at the residual-mean-square (RMS) calculated from the residual (close-loop; CL) reconstructed wavefronts. Note we use the RMS as an intermediate performance metric and then use the measured contrast as our final metric. We make use of two different CL wavefront reconstructed values: filtered and full CL wavefront. For both cases, the CL values (i.e., residuals) are calculated in DM units of voltages and then a conversion factor of $0.6\frac{\mu m}{V} $ is applied to convert between DM voltages and optical-path-difference/wavefront. 

The filtered CL DM commands are those reconstructed in real-time by the RTC from the residual/CL PyWFS slopes. Modal filtering is embedded in the CM and so the resulting residual is filtered depending on the applied modal gains and mode cut-off for the leaky integrator. In post-processing the saved AO telemetry is used to reconstruct the full CL DM commands using the full command matrix without any modal filtering. The filtered and full wavefront residuals are then computed from their respective CL DM commands using the conversion factor and finally the filtered and full RMS wavefront errors are computed represented by $CL_{filt}=CM \times [S(t)-S_{ref}]*0.6$ and $CL_{full}=CM_{full} \times [S(t)-S_{ref}]*0.6$, respectively.


\section{Daytime testing}
\label{sec:daytime}
As outlined in Sec.~\ref{sec:controller} changes were made to the current predictive filter implementation and new features such as RLMMSE were added. In this section, we present results from our daytime tests on the Keck II AO bench including results from these new features that have been tested during the day. 

\begin{figure}[ht]
    \centering
    \includegraphics[width=\textwidth,trim={0cm 0cm 0cm 11cm},clip]{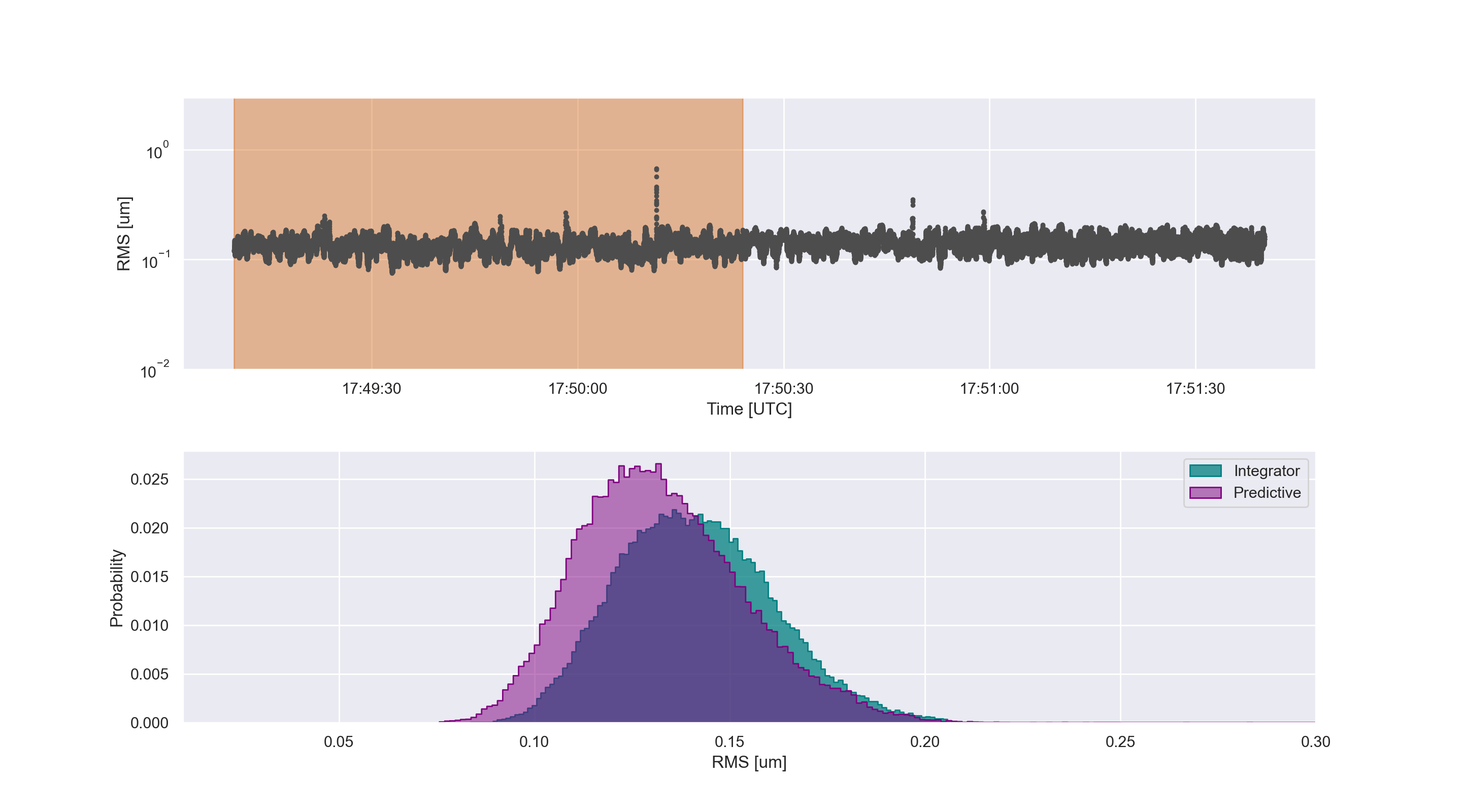}
    \caption{The plot shows the histogram of the RMS wavefront error for the predictor and integrator from data taken during the day on May 24 2021 while using the RLMMSE algorithm. The probability is the counts per bin normalized to the total counts in each dataset such that the sum of all the bins in each histogram is 1. The predictor's median RMS value is $1.1~\times$ smaller than that of the integrator.  These data were taken while playing single layer turbulence on the DM with $r_0 =16~cm$ and a wind speed of 15~$m/s$. }
    \label{fig:timeseries_RLMMSE}
\end{figure}
In Fig.~\ref{fig:timeseries_RLMMSE}, we show the results from testing the RLMMSE algorithm with single layer turbulence imposed on the Keck II DM on May 24, 2021. The filtered RMS plotted is the residual wavefront error as measured by the PyWFS. The modest improvement shown in Fig.~\ref{fig:timeseries_RLMMSE} confirms that the implementation and the algorithm work as expected. \editsA{The modest improvement is aligned with the results from Doelman (2019)~\cite{Doelman_2019}. Operating at 1kHz, the turbulence moves less than 1 sub-aperture each frame. Hence, it is more relevant to think of each sub-aperture as a time series than to consider the spatial correlations between sub-apertures as done using Frozen Flow Hypothesis.} More day time testing is needed to find the optimal parameters for the RLMMSE (such as the initial training time; data was taken with a few thousands frames of training, much less than EOF) as well as to improve the rate at which the RLMMSE updates. Currently, updating the filter is done within python and takes around 0.02 seconds; ideally the algorithm will be implemented in C~\editsA{ enabling much faster update rates. With the current implementation, however, we see a stable performance indicating that the statistics of the system are not changing faster than our update rate with the AO loop running at 1kHz.}  

\begin{figure}[ht]
    \centering
    \includegraphics[width=\textwidth,trim={0 0cm 0cm 0cm},clip]{ 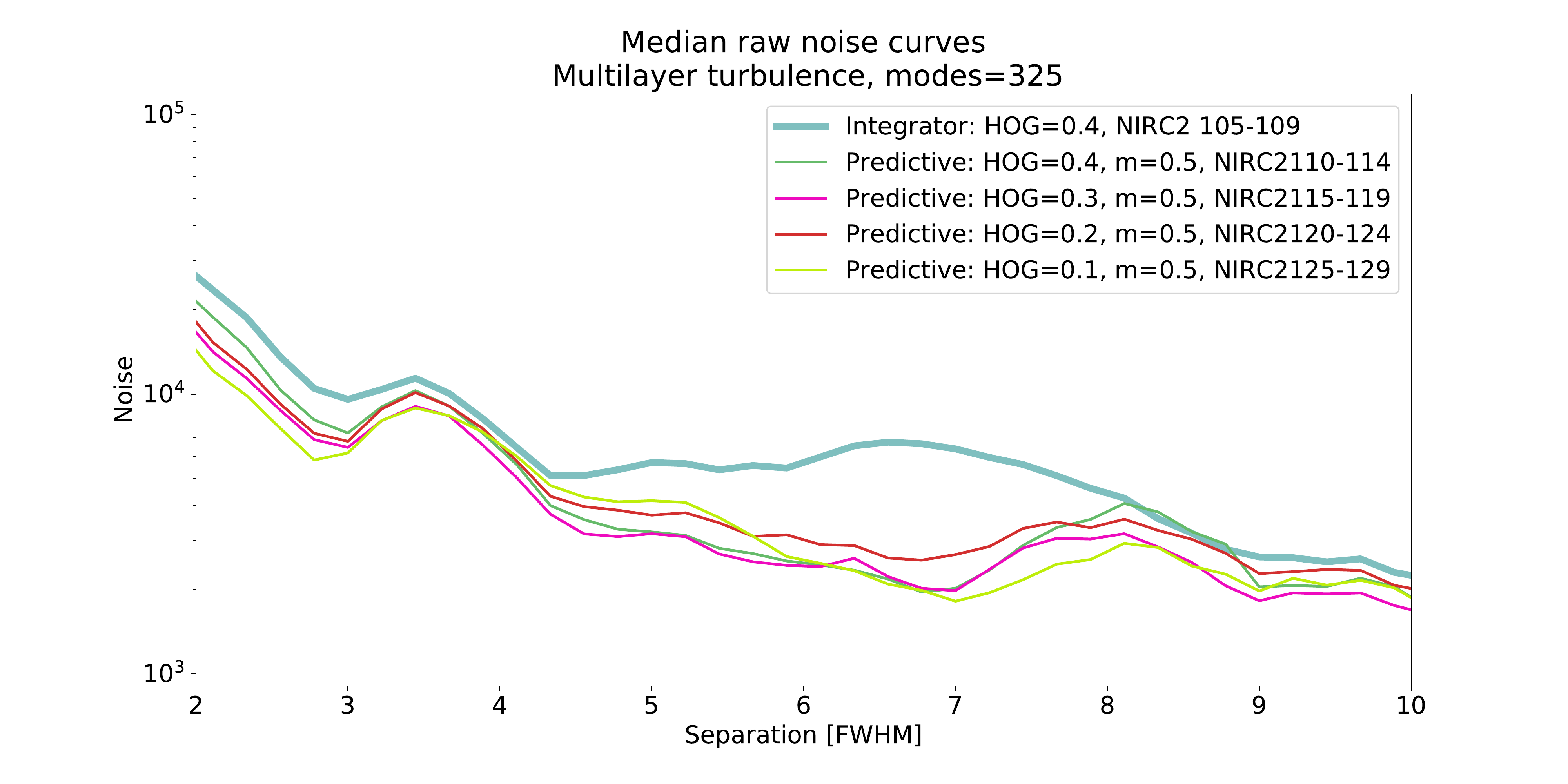}
    \caption{ Daytime median noise curves \editsA{in counts} for K-band coronagraphic NIRC2 images taken with 325 controllable modes~\editsA{taken on July 27, 2021}. The integrator gain was set to 0.4 during the integrator dataset - the standard value for similar conditions. During the predictive dataset, the integrator gain was reduced to 0.2. The mixing factor ($m$) varied slightly while we changed the HO gain. The applied atmospheric turbulence on the DM was a multi-layer atmosphere model~\cite{KAON303} with piston, tip, and tilt removed and a $r_0$ of 20~cm.}
    \label{fig:daytime_contrast_modes325}
\end{figure}

\begin{figure}[ht]
    \centering
    \includegraphics[width=\textwidth,trim={0 0cm 0cm 0cm},clip]{ 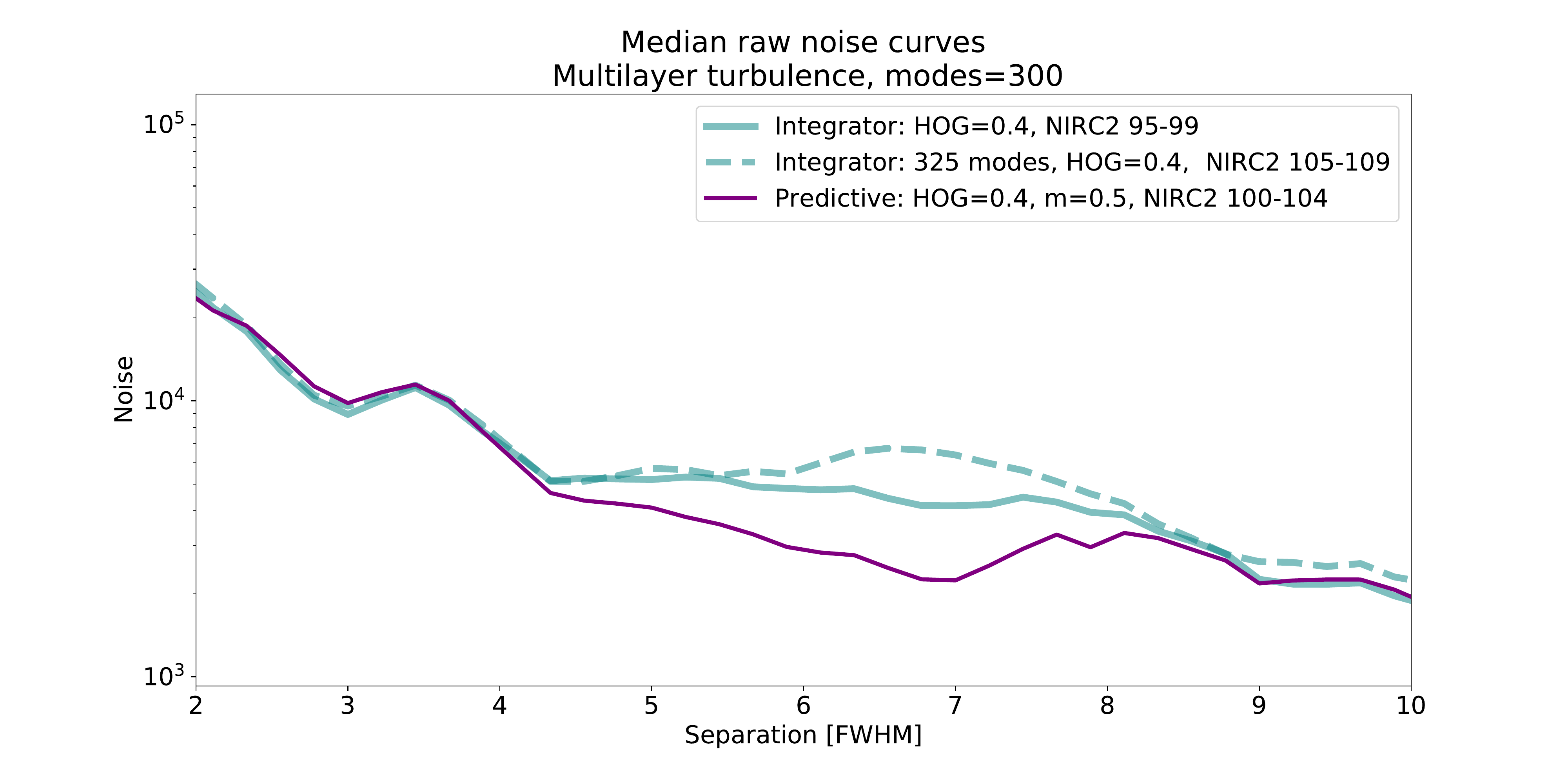}
    \caption{ Daytime median noise curves \editsA{in counts} for K-band coronagraphic NIRC2 images taken with 300 controllable modes~\editsA{taken on July 27, 2021}. The integrator gain was set to 0.4 during the integrator dataset and then reduced to 0.2 for the predictor. The applied atmospheric turbulence on the DM was a multi-layer atmosphere~\cite{KAON303} model with piston, tip, and tilt removed and a $r_0$ of 20~cm. \editsA{The dashed integrator is the same curve as shown in Fig~\ref{fig:daytime_contrast_modes325}}.}
    \label{fig:daytime_contrast_modes300}
\end{figure}

During the day of July 27, 2021, we tested EOF under a variety of different configurations and focused on varying the number of modes and the HO gain while taking K-band coronagraphic images with NIRC2. We are limited to K-band by the internal light source, however, for on-sky tests we make use of the L-band vortex. We applied multi-layer turbulence with the parameters outlined in the Keck Adaptive Optics Note No 303 (KAON303)~\cite{KAON303} removing piston, tip, and tilt. Setting $\alpha=1$, and training on 2 minutes of data, the EOF provided an improvement over the nominal integrator configuration, as described below.  We plot the median raw noise curves for 325 controlled modes in Fig.~\ref{fig:daytime_contrast_modes325} and for 300 controlled modes in Fig.~\ref{fig:daytime_contrast_modes300} of the NIRC2 images throughout the daytime tests. These images show the contrast out to the DM control radius of 10$\lambda/D$. The noise as a function of separation was computed using the Vortex Image Processing (VIP~\cite{2017AJ....154....7G}) python package: a ring of FWHM-diameter circular apertures was constructed at each separation, and the noise was taken to be the standard deviation of the aperture sums at a given separation.  Figures.~\ref{fig:daytime_contrast_modes325} and \ref{fig:daytime_contrast_modes300} show that the predictor is able to perform better (by a factor of 3.5 for 300 modes and 4-5 for 325 modes) than the integrator at small separations (4-7 $\lambda/D$; where we expect predictive control to provide an improvement) under a variety of different system configurations. Figure~\ref{fig:daytime_contrast_modes325} shows that varying the HO gain for the predictor does not have a large impact on its performance. We, however, acknowledge that this behavior might be very different on-sky because for this experiment we are introducing turbulence using the DM and therefore we do not have the same amount of HO spatial frequencies \editsA{of atmospheric turbulence that would cause aliasing on the PyWFS that }we would on-sky. ~\editsA{Finally, comparing the daytime integrator's performance, we see that the integrator is more affected by the number of corrected modes than the predictor and that fewer modes provide a better noise curve (comparing the two integrators plotted in Fig~\ref{fig:daytime_contrast_modes300}). This might be due to our Zernike modes being defined on a circular aperture but applied to a segmented pupil or that this is the response of our imaging arm with the vortex coronagraph.} 

\section{On-sky testing}
\label{sec:nighttime}

In this section, we present the results from our two engineering nights and one science night in 2021 that allowed for on-sky testing of predictive control. 

\subsection{Night of June 20, 2021}
We were able to close the AO loop on J21095901+2859392 - a bright star with an H-band magnitude of 4.24. Issues with the secondary mirror of the telescope and the suboptimal performance of the AO system resulted in low SR and the decision to take non-cornographic images. These issues mean the results from June 20, 2021 cannot be used conclusively to demonstrate the performance of predictive control compared to the integrator. However, the night allowed us to verify the functionality of our implementation and observe the overall behavior of the predictive controller. We note that during our actual tests no seeing data was available from the Canada-France-Hawaii Telescope's (CFHT) MASS and DIMM instruments~\cite{CFHT_weather}. 

\begin{figure}[ht]
    \centering
    \includegraphics[width=0.95\textwidth,trim={0 1.5cm 1cm 0cm},clip]
    {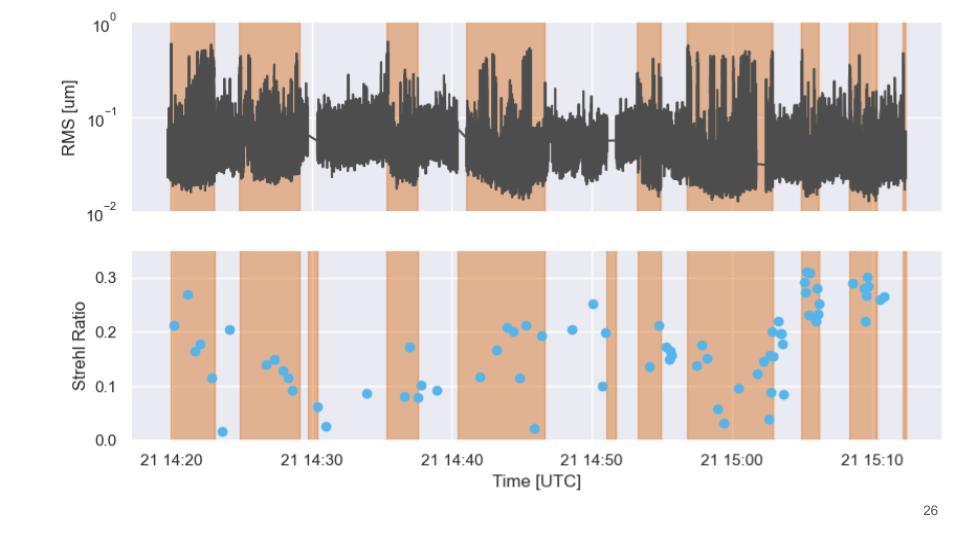}
    \caption{The top panel shows a time series RMS wavefront error ($CL_{filt}$) for the entire testing period taken starting the night of June 20, 2021. \editsA{Throughout the night different parameters are changed including the lag and the gain, which is changed continuously during prediction to maximize (although it sometimes degrades) the performance of the predictor. Looking at the minimum RMS values will help the reader to see the performance differences as the spikes occur from telescope offloading and DM saturation.} The bottom panel shows the Strehl ratio as measured from the NIRC2 images~\editsA{in Brackett Gamma}. The shaded orange region indicates when prediction was turned on. }
    \label{fig:timeseries_June20}
\end{figure}

\begin{figure}
    \centering
    \includegraphics[width=0.85\textwidth,trim={5.5cm 3.cm 5.5cm 3.4cm},clip]{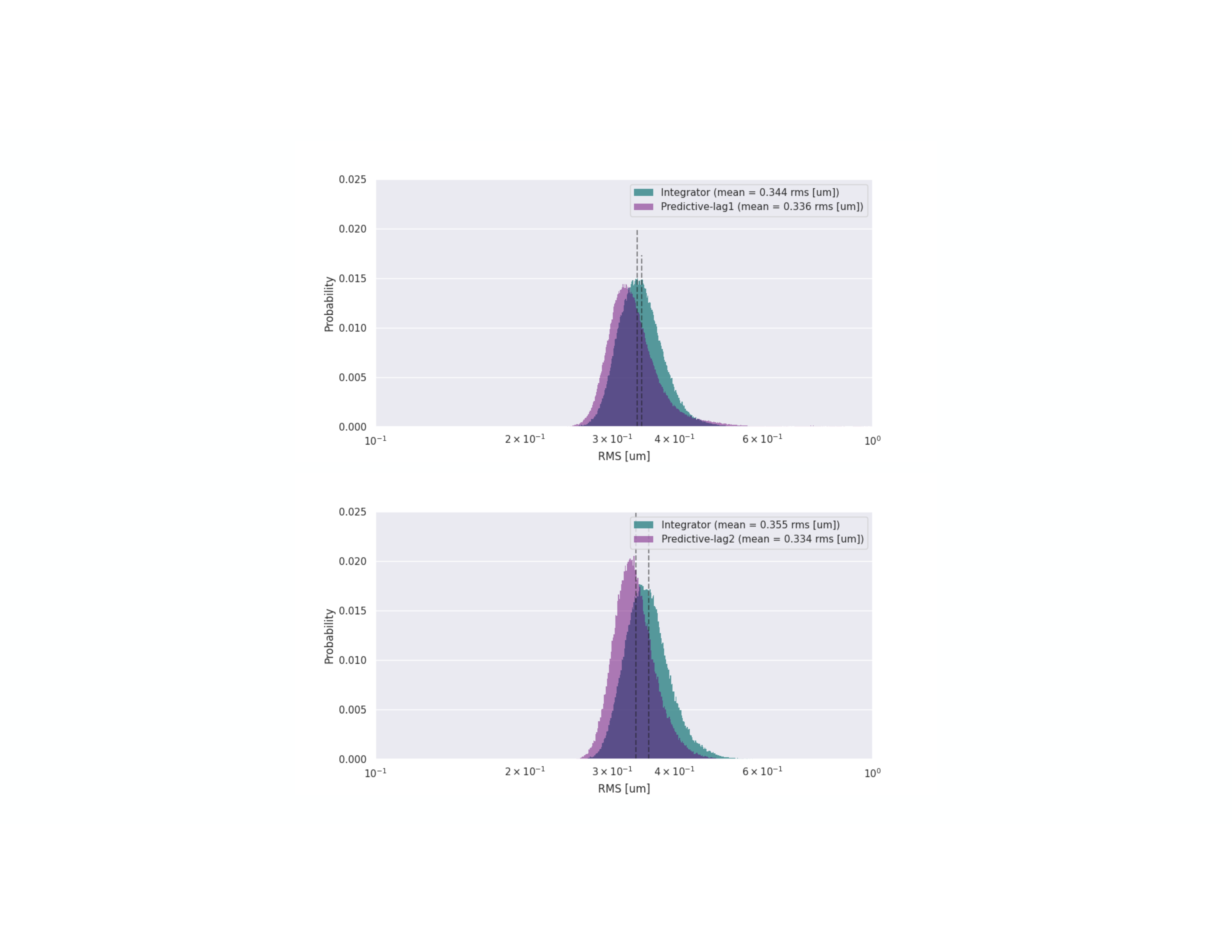}
  
   \caption{Analysis of night time tests on June 20, 2021.  Top panel shows for 1 frame lag and bottom panel shows data for 2 frame lag predictor compared to the integrator.  The integrator data was taken before and after prediction was turned on and off.  The probability is the counts per bin normalized to the total counts in each dataset such that the sum of all the bins in each histogram is 1. }

    \label{fig:hist_June20}  
\end{figure}

\begin{figure}
    \centering
    \includegraphics[width=0.75\textwidth,trim={0cm 0cm 0cm 0cm},clip]{ 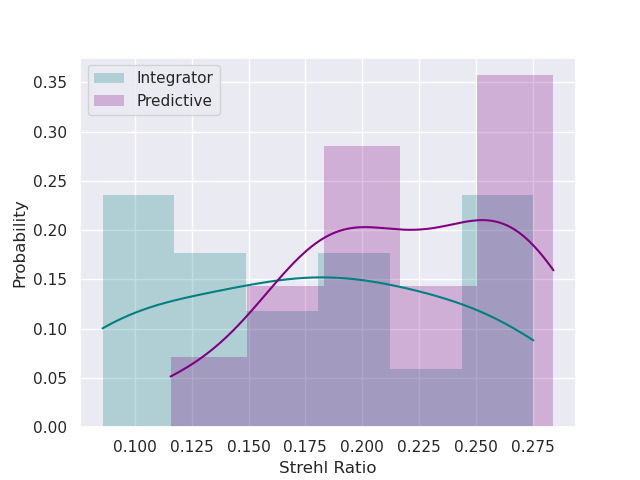}
    \caption{The histogram shows the SR~\editsA{in Brackett Gamma} for predictive control with a lag of 2 and for the leaky integrator~\editsA{for data taken on June 20, 2021}. The solid lines indicate the kernel densities for the data. }
    \label{fig:SR_jun20}
\end{figure}

In Fig.~\ref{fig:timeseries_June20}, we show the filtered RMS wavefront error (i.e., $CL_{filt}$) for the entire night. We also show the SR as measured from NIRC2 images. Studying the time series of the RMS, we can see the improvements provided by prediction as the RMS becomes smaller when predictive control is turned on (shaded orange regions; \editsA{it is helpful to track the minimum RMS values to see the performance difference and not the maximum RMS values that are contaminated by telescope offloading, DM saturation, and other effects}). The time series also reveals the effects of all the changes made to the AO controller while observing. The continued improvement of prediction in a shaded region (i.e., between times 14:40 and 14:45) is a result of the mixing factor and integrator gain being adjusted to find the optimal configuration. During this observation period, we also tested the effects of changing the assumed frame lag in the POL step. Figure~\ref{fig:hist_June20} shows the results from a lag of 1 and lag of 2. With a lag of 1 we are able to improve the performance, but the histogram is skewed with a tail tending towards larger RMS values. This tail is removed when a lag of 2 is used by the predictor. We also confirm the lag of 1.7 frames as determined by S. Cetre et al. 2018~\cite{Cetre_2018}. We will investigate implementing a non-integer lag in the near future for our predictive control filter for further improvement. Furthermore, the lag of 2 histogram shows a greater improvement in RMS wavefront error. Comparing the NIRC2 SR for the integrator and predictor with lag 2 in Fig.~\ref{fig:SR_jun20}, we see a clear shift towards larger SR values when predictive control is turned on. While we do find that a lag of 2 is better for the predictor and the predictor performs better than the integrator it is difficult to draw concrete conclusions on the provided improvement due to the lack of seeing data and the secondary mirror issues.


\subsection{July 27, 2021}
On the night of July 27, 2021, we received a half night of engineering time for predictive wavefront control tests~\editsA{during which the performance of the AO system was typical}. In this section, we present the results for one of our targets, HIP 117578. With an H-band magnitude of 6.1, HIP 117578 is bright enough to take advantage of the 1kHz frame rate of the PyWFS camera. We took NIRC2 images with the L-band coronagraph in place and made use of QACITS to keep the point-spread-function (PSF) centered on the vortex coronagraph~\cite{QACITS}. 

\begin{figure}
   \centering
    \includegraphics[width=\textwidth,trim={15.5cm 0cm 15.5cm 0cm},clip]{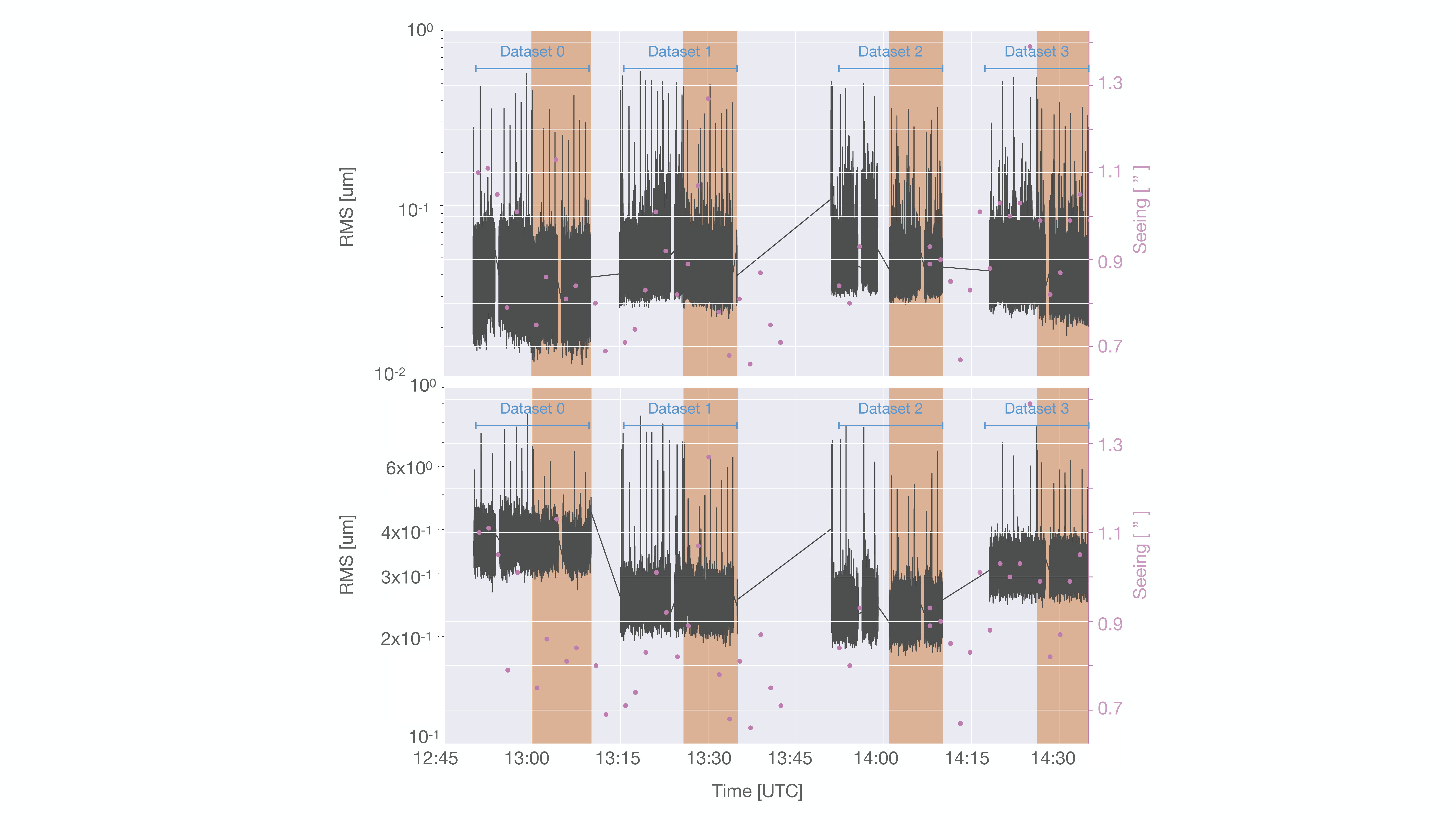}
    \caption{$CL_{full}$ (top) and $CL_{filt}$ (bottom) time series for observations~\editsA{for data taken on July 27, 2021} of HIP 117578 with the shaded regions indicating when prediction was turned on. The seeing in both panels is the DIMM as measured by CFHT~\cite{CFHT_weather}. We indicate four different datasets that were taken with various configurations as outlined in Tab.~\ref{tab:July28_parameters}.}
    \label{fig:timeseries_July27}
\end{figure}

\begin{figure}
    \centering
\includegraphics[width=1\textwidth,trim={3cm 2.5cm 3cm 2.5cm},clip]{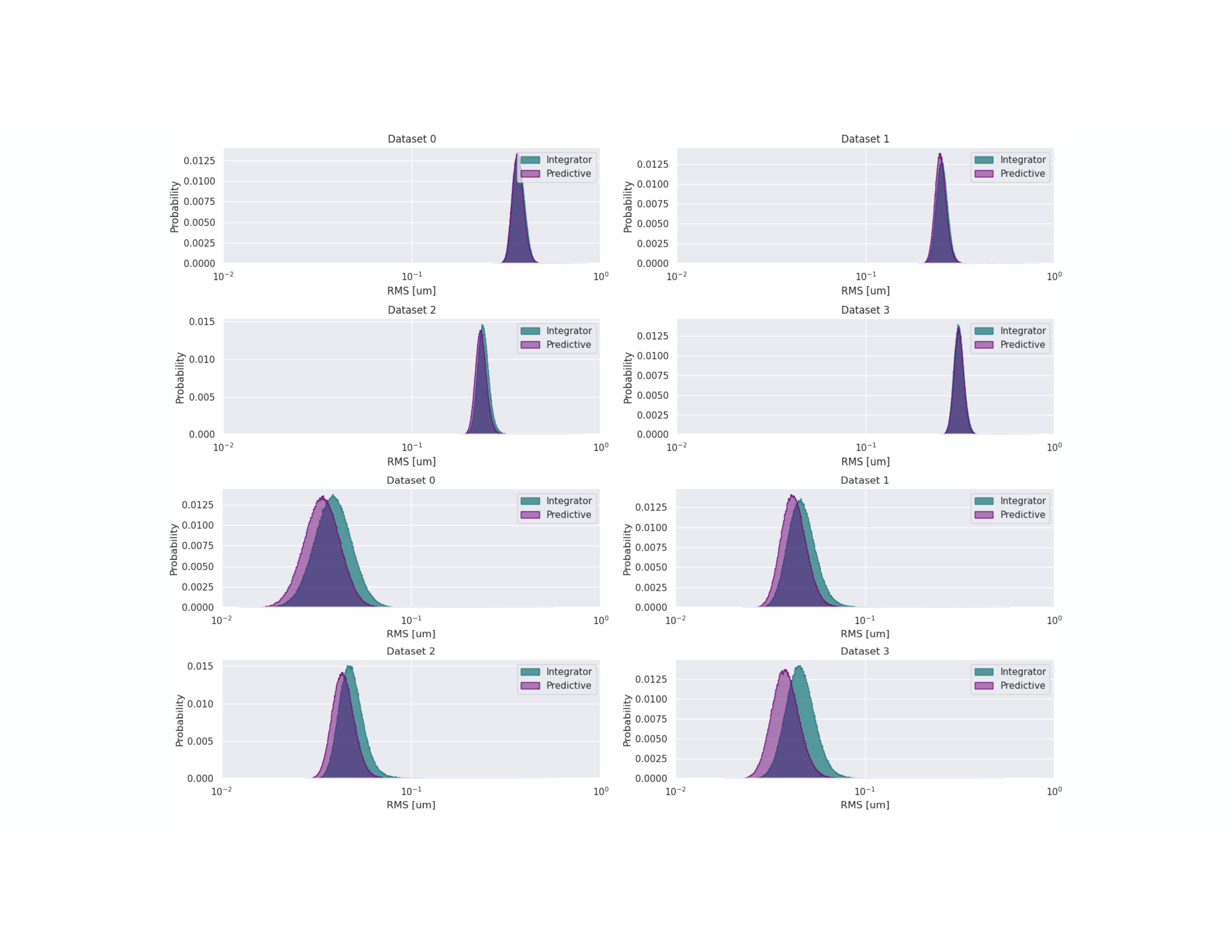}

   \caption{Probability (i.e., counts normalized by total number of data points) for each of the different data sets separated by controller type. We compare the predictor with the integrator data taken just before the predictor is turned on. The different parameters for the data sets~\editsA{taken on the night of July 27, 2021} are provided in Tab.~\ref{tab:July28_parameters}. }

    \label{fig:hist_July28}  
\end{figure}

Figure~\ref{fig:timeseries_July27} provides an overview of the observations with the RMS wavefront error being plotted as a function of time for both the full and filtered RMS values (i.e., $CL_{full}$ and $CL_{filt}$). We also plot the seeing as measured by the DIMM from CFHT. The figure indicates four data sets that were taken for this target. For each data set we configured the AO system as desired (see Tab.~\ref{tab:July28_parameters}) and then started the QACITS sequence which first runs an optimization sequence to center the vortex and then takes a pre-determined number of science images. Half way through this set of science images we switch from the integrator controller to the predictive controller. These full QACITS sequences, which include NIRC2 frames taken using the integrator followed by predictive control, constitute a ``dataset" (Tab.~\ref{tab:July28_parameters}). The NIRC2 images were all taken with an exposure time of 0.3 seconds and 90 coadds at the full frame of 1024-by-1024 pixels. When we swap controllers, one to two NIRC2 frames are contaminated and hence dropped from this analysis. Therefore, we compare the predictor to integrator data that was taken just before the predictor data set. In Fig.~\ref{fig:hist_July28}, we plot the histogram of the RMS wavefront error for the four different data sets once again for the filtered and full RMS values. These histograms only include data that were taken during the NIRC2 exposures that are used for our contrast analysis. The ratio of the medians and standard deviations (std) of the histogram for each dataset are presented in Tab.~\ref{tab:July28_parameters} for the filtered and full RMS values. From these values, we see that the predictor provides an improvement in median wavefront error for all cases when looking at the filtered wavefront error which makes sense as the LMMSE aims to minimize the residual wavefront error and we trained on the filtered wavefront. For the full wavefront, the predictor's median is worse than that of the integrator for dataset 3. The standard deviation for dataset 2 using both the filtered and full residuals wavefronts is improved by the predictor which indicates that the predictor provided a more stable AO correction than the integrator during this time. 

\begin{figure}
    \centering
   \includegraphics[width=0.95\textwidth,trim={1cm 2.5cm 1cm 2.5cm},clip]{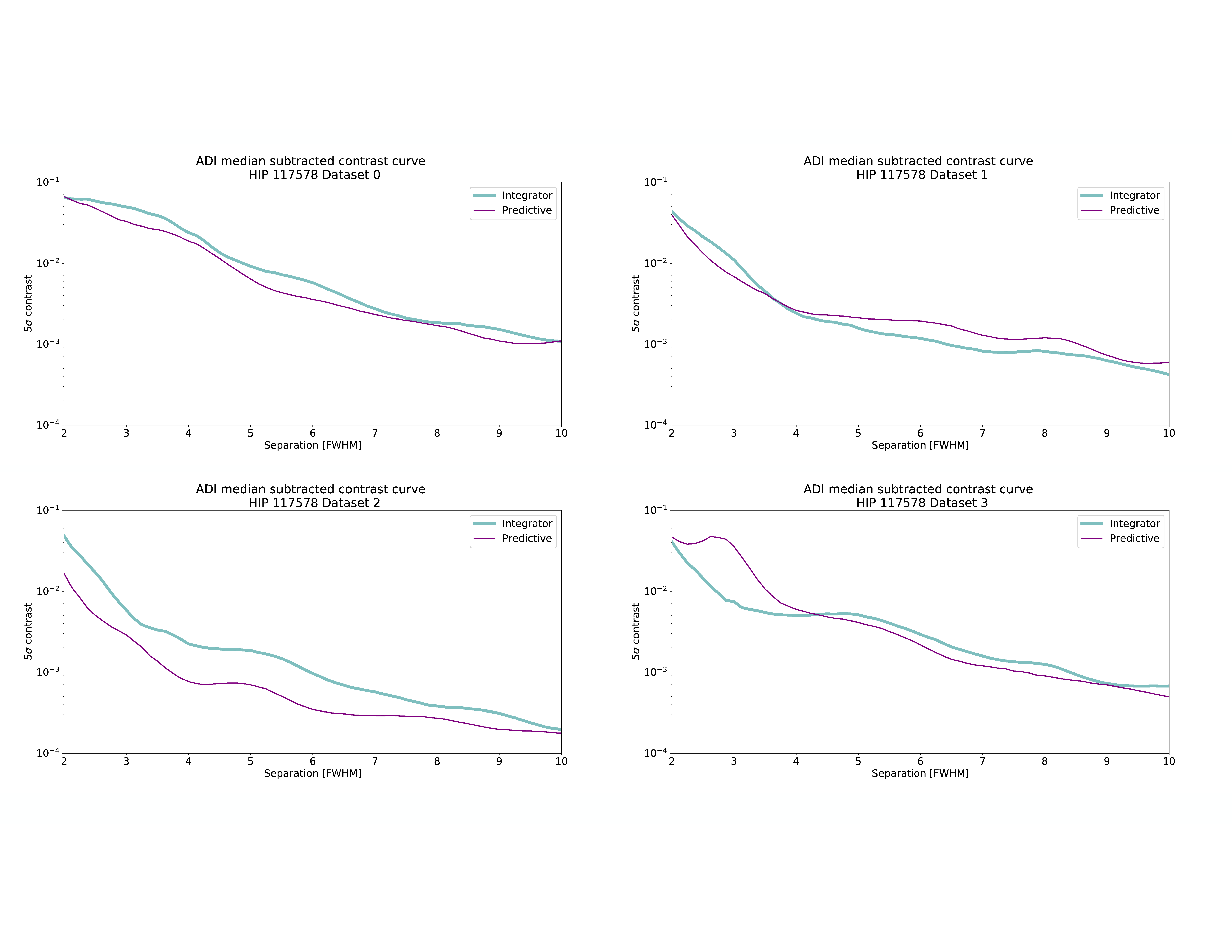}

   \caption{Angular differential imaging contrast curves for HIP 117578 taken on the night of July 27, 2021 for the four different datasets as outline in Tab.~\ref{tab:July28_parameters}.  }

    \label{fig:contrast_july28}  
\end{figure}

\begin{figure}
    \centering
    \includegraphics[width=16cm,trim={0cm 8cm 0cm 5cm},clip]{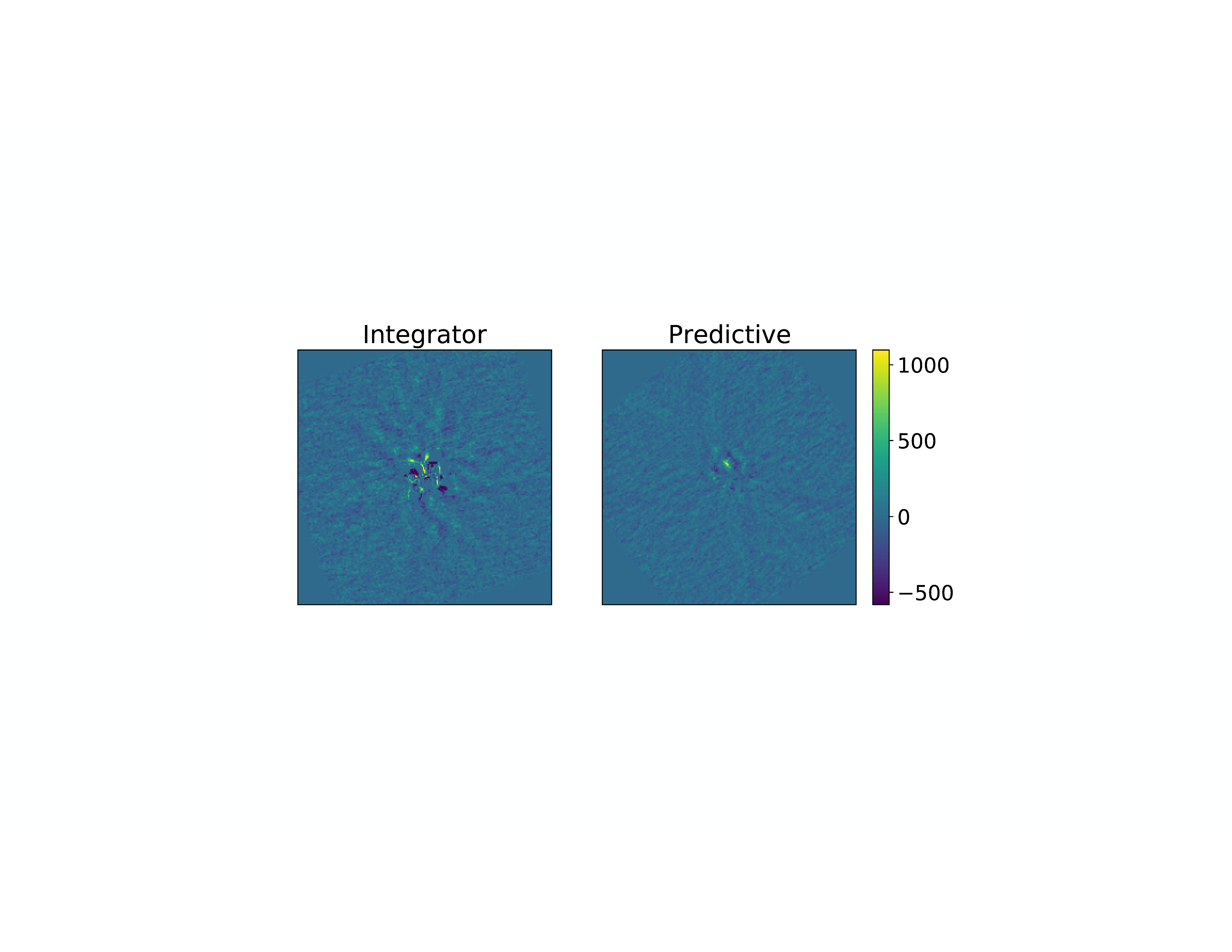}
    \caption{\editsA{Post-processed coronagraphic PSFs for the integrator and predictor corresponding to the contrast curves in Fig.~\ref{fig:contrast_july28} for dataset 2 taken the night of July 27, 2021 for HIP 117578. }}
    \label{fig:psf_july28}  
\end{figure}

We present the contrast curves for the four different data sets taken while observing HIP 117578 in Fig.~\ref{fig:contrast_july28}. The NIRC2 data was first bad pixel, flat-field, and sky corrected, as well as registered, using the automated pipeline described in Xuan et al.~2018\cite{2018AJ....156..156X}. We used VIP to de-rotate, median subtract, and median combine the data, modifying the number of images in each dataset such that the integrator and predictive components had the same total parallactic rotation. We then used VIP to create student-t corrected, algorithmic throughput corrected contrast curves, as detailed in Gomez Gonzalez et al.~2017\cite{2017AJ....154....7G}.  From the contrast curves we find a contrast gain of 2 for dataset 2 at a separation of 3~$\lambda/D$ as reported in Tab.~\ref{tab:July28_parameters}. In all of the datasets, however, we see an improvement in contrast located at different separations, which can be seen in Fig.~\ref{fig:contrast_july28}. Figure~\ref{fig:contrast_gains_july28} in App.~\ref{app:gains} shows the contrast gain (ratio of the integrator and predictor contrast curves) for the four data sets, clearly showing the achieved improvement with predictive control. In data set 2 we see contrast gains close to 3 at separations of 4-6 $\lambda/D$ while we see more modest gains around 1.5 for data sets 0 and 3. From these contrast curves, we have repeatable improvement in contrast with EOF predictive control take over a period of 1.5 hours while the only time we do not have a contrast improvement (dataset 3) we made a significant change to order of the predictive filter which we had not previously tested. \editsA{We also plot the post-processed corongraphic PSFs for dataset 4 in Fig~\ref{fig:psf_july28}, further illustrating the spatial improvement provided by the predictor. }

Studying all the data, \editsA{it} appears that the ratio of the full std of the RMS wavefront error is the best indicator for contrast improvement \editsA{as shown in Tab~\ref{tab:July28_parameters}}. When the full std ratio is less than 1 the contrast is worse, when it is equal to 1 we have slight improvement and for larger than 1 we have the greatest contrast improvement. 

\subsection{October 25 2021}
\begin{figure}
    \centering
        \includegraphics[width=10cm, trim={1cm 1cm 3cm 0.5cm},clip,page=1]{ 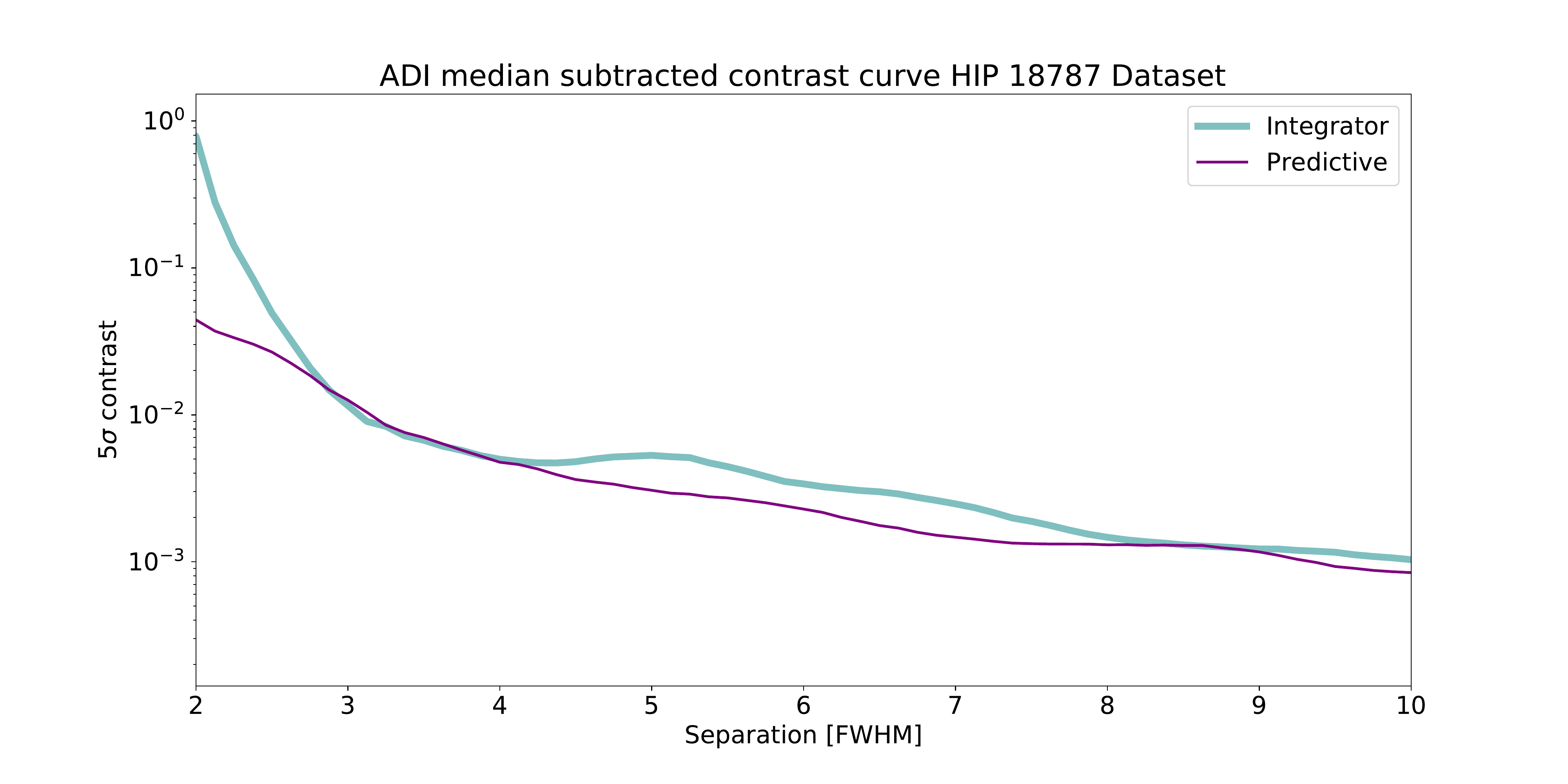}
   \caption{Angular differential imaging contrast curves for HIP25645 taken on the night of Oct 25, 2021.  }

    \label{fig:contrast_oct25}  
\end{figure}

We did more testing with EOF predictive control during science observations on the night of Oct 25, 2021~\editsA{for which we had typical AO performance}. Not only were we able to demonstrate an improvement in residual wavefront error again but also an improvement in contrast. Figure~\ref{fig:contrast_oct25} shows the ADI median subtracted contrast curve for HIP 25645 which has a H-band magnitude of 6.77. The contrast is improved between 5-8 $\lambda /D$ which is in agreement with the results from July and also close between 2-3$\lambda /D$. For these observations we corrected 212 modes with a HO and integrator gain of 0.28 and 0.4, respectively. These parameters were set by Keck personnel following standard AO operations.  A mixing factor of 0.5 and gain of 0.23 where used for predictive control. 

\subsection{Summary of controller performances}
\begin{figure}
    \centering
        \includegraphics[width=0.85\textwidth, trim={0cm 0cm 0cm 0cm},clip,page=1]{ 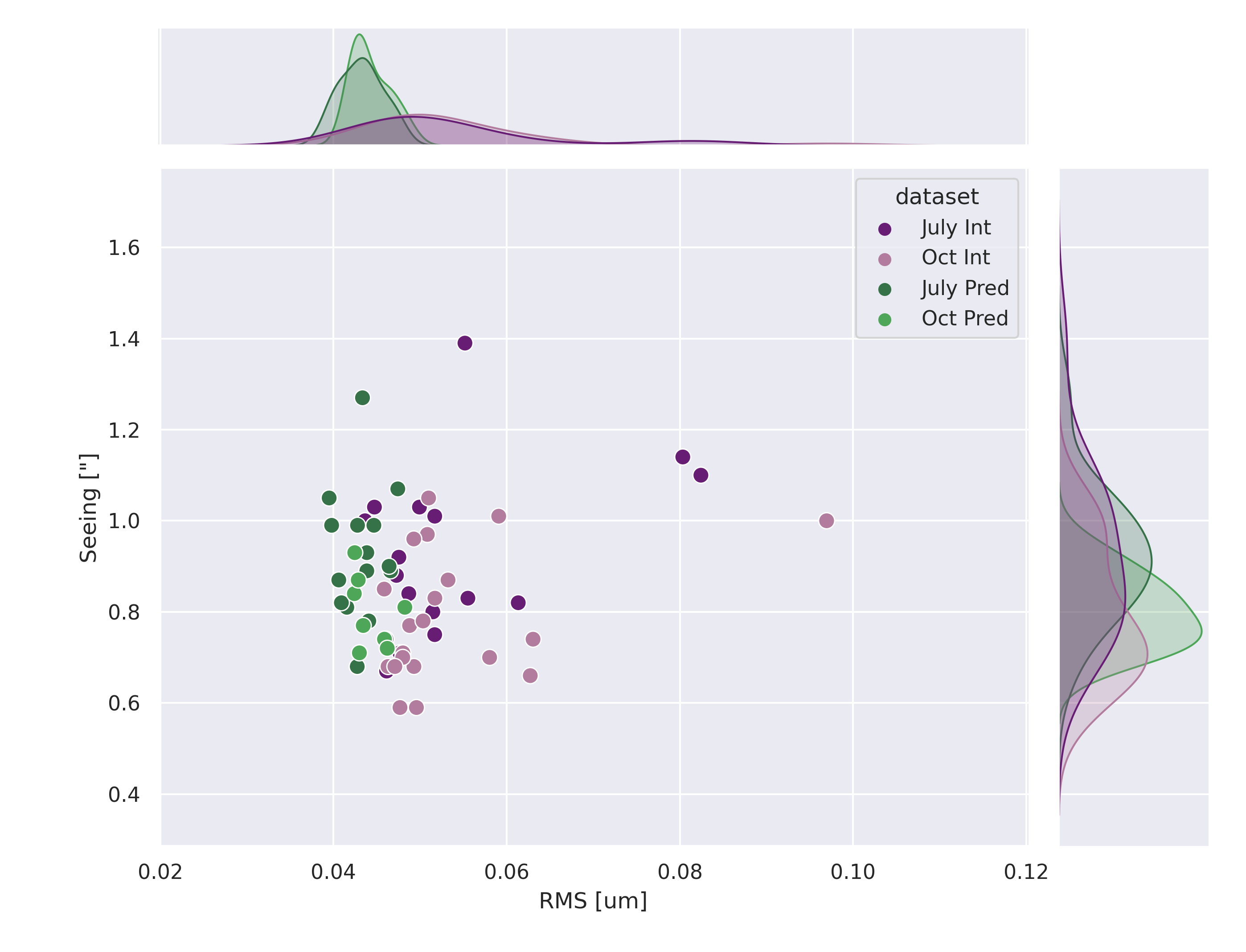}
   \caption{The seeing and $CL_{filt}$ for the predictor and integrator data for the two nights of~\editsA{July 27 and October 25 2021}. The side panels show the normalized kernel density plots for the seeing (right) and the $CL_{filt}$ (top). }

    \label{fig:JO_all}  
\end{figure}

The RMS wavefront error results from our latter two nights (Jul and Oct~\editsA{for which we had good AO performance}) are summarized in Fig~\ref{fig:JO_all}. We plot the seeing against the filtered residual RMS wavefront error. We take the CFHT seeing value and resample the RMS wavefront error to every 10 seconds to get a mean correction and then take the mean RMS wavefront values within 30 seconds of the CFHT seeing value. We make use of all the available predictor data that was taken. The upper and side subplots show the kernel density functions. The plots show that distribution of seeing values for all the data is roughly the same for both controllers (with the seeing actually being better in some cases for the October integrator than the predictor) on the two nights while the distribution in residual wavefront error is different for the two controllers over both nights. This indicates that the difference in the predictor and integrator is not due to a change in seeing. The predictor on both nights shows a consistent reduction in wavefront error compared to the leaky integrator as illustrated by the scatter plots being shifted to the left for the predictor datasets compared to the two integrator datasets. There is also a clear reduction in the spread of wavefront errors for predictive control showing that for different conditions the predictive controller has similar performance resulting in a more reliable and consistent performance. This is in agreement with what was found by van Kooten et al 2020~\cite{vanKooten_2020}. 

\section{Discussion} 
\label{sec:discussion}
\subsection{Performance metric}
We compare, in this paper, the residual wavefront error and raw contrast of predictive control to the current controller performance at Keck, the leaky integrator. More specifically, we define an improvement compared to the integrator as setup by Keck personnel such as the telescope operator, observing assistant, or staff astronomer who follow guidelines to provide robust, good AO performance depending on the conditions. This protocol was strictly followed for the Oct 25 night while loosely followed for observations in June and July (e.g., we manually increased the number of modes such that the integrator had the best performance). During June and July we altered the leaky integrator to perform even better than what was suggested by the operator by mainly correcting for more modes than recommended. In the end, our implementation of predictive control on Keck aims to produce the best performance possible with the current system through a data-drive approach. Therefore, our chosen metric is the current performance of the system. 

\editsA{Under the assumption that the servo-lag error is a large contributor to the wavefront error and dominating the contrast at small angular separations on NIRC2~\cite{2018AJ....156..156X}, we implemented a data-driven predictor, EOF, to reduce the temporal wavefront error. Since the real goal is to improve the contrast at small angular separations, we, therefore, attribute an improvement in contrast to a reduction in the servo-lag error. Our method, however, relies on both spatial and temporal correlations, therefore we are also able to correct for other sources of errors in the system (i.e., calibration and registration alignment), making this an attractive method in general for improving the overall performance of an AO system. }

\subsection{Discussion of results}
With the our predictive filter in place and demonstration of contrast improvement over a couple of nights and reduction in RMS wavefront error over several nights, we can transition towards making an comprehensive review of optimum parameters (see Sec.~\ref{sec:future}). 

We began such testing in July 2021 where we aggressively varied the HO gain between 0.3 and 0, where 0 value indicates that we are not controlling any HO modes. It is when we are not controlling the HO gains (dataset 2) that we achieved the largest improvement in contrast. Comparing dataset 1 and 2 where the HO gain goes from 0.3 to 0, the contrast curves in Fig.~\ref{fig:contrast_july28} show the integrator is able to perform slightly better at larger separations than the predictor due to HO modes that are being controlled. Comparing then to the data taken in Oct shown in Fig.~\ref{fig:contrast_oct25}, where the HO gain value is similar to dataset 1, we are able to perform better with the predictor. We note that the number of total corrected modes are less in Oct as the number of modes needed to be lowered to provide a stable correction for the leaky integrator. Therefore, at larger separations, it is difficult to know what HO and number modes to use for the best improvement with prediction as it is clearly driven by external factors such as the atmospheric conditions. Close-in to the star, however, the predictor performs better regardless of the HO gains in those datasets. The transition separation between these two regimes appears to be around 4~$\lambda/D$. 

Finally, we point out the behavior of the filtered and full residual wavefront error as shown in Fig.~\ref{fig:timeseries_July27}. In the bottom panel we plot $CL_{filt}$ which is used to generate the POL wavefront error that is used to generate the predictive filter by minimizing the mean square error. As we change various parameter, the $CL_{filt}$ varies in clear explainable way where for example by increasing the $g$ from dataset 0 and dataset 1 decreases the residual wavefront error. In the top panel of Fig.~\ref{fig:timeseries_July27} we plot $CL_{full}$ where we include the HO modes that we are not controlling into the residual wavefront error. Between datasets it varies in the opposite way from $CL_{filt}$. Again taking dataset 0 and dataset 1, we are increasing the gain along with the HO gain such that we are improving the residual wavefront error for the controlled modes. We, however, appear to be increasing the full residual wavefront error. Testing along with end-to-end simulations are needed to better understand this behavior starting for the leaky integrator.

\subsection{Power spectral density analysis}

\begin{figure}
    \centering
    \includegraphics[width=0.8\textwidth,trim={0cm 0cm 0cm 0cm},clip]{ 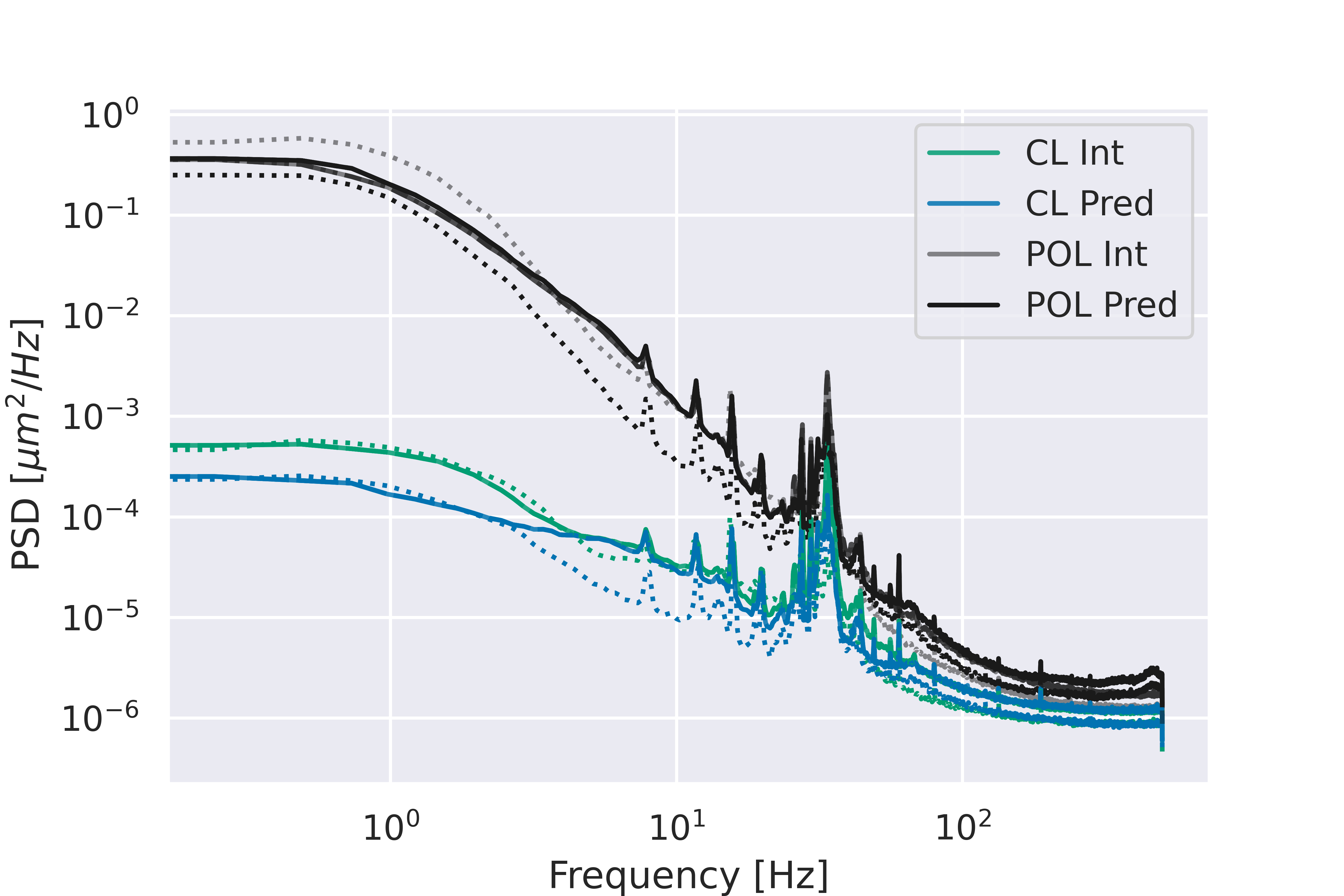}
    \caption{Power spectral densities calculated using the Welch method for the $CL_{filt}$ (indicated by CL) and the corresponding pseudo-open loop (POL) integrator and predictor data sets. \editsA{The four different line styles correspond to the different data sets~\editsA{taken on the night of July 27, 2021 and} outlined in Tab.~\ref{tab:July28_parameters} with the solid line corresponding to data set 2 (which agrees with two other data sets perfectly)}. }
    \label{fig:Star2_PSD_July27}
\end{figure}

In Fig.~\ref{fig:Star2_PSD_July27} we present the estimated power spectral densities (PSDs) for the on-sky July data. From the POL PSDs we see that temporally the optical turbulence did not vary significantly between datasets and swapping of the controller. In closed-loop, the predictor reduces the wavefront error at small frequencies while also reducing the peak near 60 Hz. While the predictor PSD is more flat compared to the integrator, it also indicates that a greater improvement in performance is still possible by removing the peak around 60 Hz.  
To further improve the system and the performance of the AO system, we must understand where different features of the PSDs originate from. In the CL PSD the peak at near 60Hz has the same amplitude as the low frequencies suggesting that this peak is now a main source of the residual wavefront error. The frequency of the peak is similar to one seen in a tip/tilt PSD indicating that both peaks might have the same source but further investigation is needed. Since EOF trains on the POL, the peak is a weak signal compared to the turbulence \editsA{such that increasing the temporal order does not improve the performance}. Therefore applying EOF to the closed-loop coefficients might help to remove this feature. We outline the steps necessary in Sec~\ref{sec:future}. 

\subsection{Comparing EOF to other controllers}
We outlined above our performance metric used in this paper. Understand the performance of predictive control in comparison to other proposed controllers is outside the scope of this paper. We instead outline work already underway and provide comments on EOF and RLMMSE as well as gain optimization.

\begin{figure}
    \centering
        \includegraphics[width=0.85\textwidth, trim={0cm 0cm 0cm 0cm},clip,page=1]{ 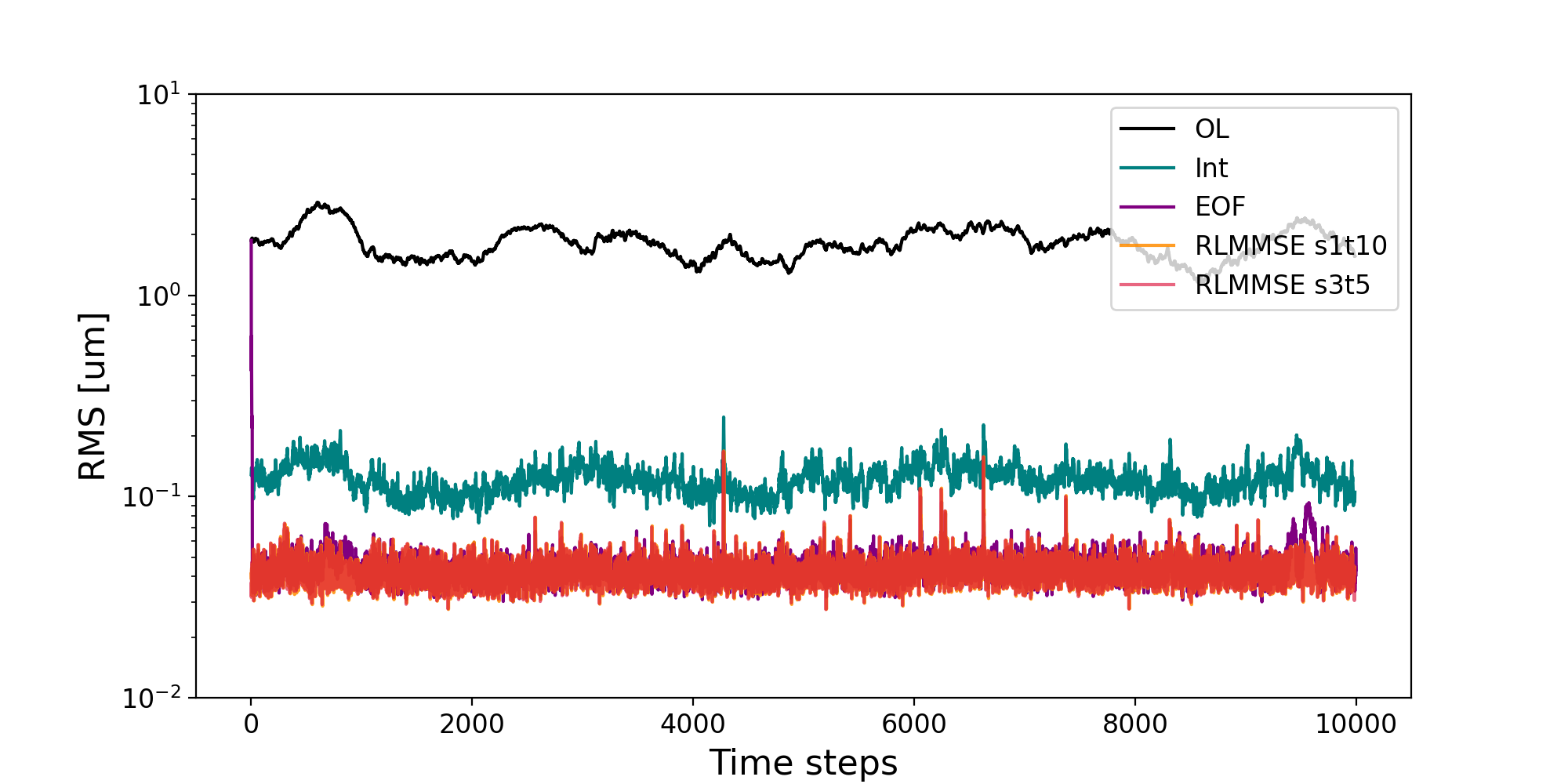}
   \caption{EOF on 30 seconds of Keck AO telemetry compared to two version of RLMMSE that use no spatial regressors and have the same temporal order (s1t10) and another which has 5 temporal orders and spatial order of a 3x3 grid around each actuator so that it correlates to its neighbors (s3t5). With EOF we train the data in advance on an earlier 30 seconds of the data while the RLMMSE trains online. The RLMMSE has many fewer regressors allowing it to converge quickly. We also plot the true open-loop (OL) and the Keck integrator (integrator).}

    \label{fig:EOF_RLMMSE}  
\end{figure}

Comparing EOF to other controllers is underway in a laboratory setting at University of California Santa Cruz on the SEAL test bench~\cite{Jensen_2021b}. We will compare predictive control not only to other types of predictive controllers but also to other controllers such as modal gain control. In our own simulations and with telemetry data the EOF and the RLMMSE algorithms have the same performance when the appropriate regressors and training data are selected. Even with fewer regressors the RLMMSE can achieve the same performance as shown in Fig.~\ref{fig:EOF_RLMMSE} where two different RLMMSE results are shown to have the same performance as EOF. The regressors and amount of training data influence the achievable performance and contribute to over- and under- fitting to the data.  In the case where we use the most recent measurement only (i.e., temporal order 1) and a spatial order of one (i.e., temporal regressors only) the RLMMSE solution is equivalent to the optimal gain for the integrator for each actuator. If we change our basis set to modes, we will have the optimal gain for each mode. Adding more regressors, depending on the system setup, might not result in better performance and in the case where the lag is one frame, more regressors might result in over-fitting. Therefore care must be taken when setting up the predictive filter. Similarly modal gain optimization relies on pseudo-open loop reconstruction that incorporates the temporal lag. The optimal gains are found by solving a specific cost function \editsA{and while the optimal gain is not a predictive algorithm, it improves the temporal performance by maximizing the correction bandwidth}. The performance of the optimal gain optimization and a data-driven predictor might be indistinguishable especially when the lag is 1 frame~\editsA{and/or for non-stationary input disturbances}. The data, regressors, optical setup of the system, and the performance metric will all contribute to how different modal gain and predictive control methods will behave. 

\section{Future work}
\label{sec:future}
Given the contrast improvement illustrated in Fig.~\ref{fig:contrast_july28}, we have a clear path forward to improve the performance of the predictive control including: \begin{enumerate*}
\item implement predictive control in a true closed-loop configuration, 
\item determine the optimal parameters for both EOF and RLMMSE, 
\item match predictive control results to end-to-end simulations expanding on our integrator simulations that match on-sky contrast~\cite{Jensen-Clem_2021_PSI}, 
\item investigate modal implementation and direct use of slopes for prediction,
and
\item continue on-sky verification of predictive control under different conditions and guide star magnitudes. 
\end{enumerate*} 

\editsA{To address some of these topics above, we are currently developing a high fidelity model of Keck II AO bench for which we can better test and match on-sky data by having the exact infrastructure in simulation. With this simulation infrastructure, we hope to  better understand the tunable parameters for future facilitization. Already, we find that the performance is improved and more stable within seconds of swapping to predictive control and little tuning is needed with selecting the nominal parameters presented in this work.}

Finally, we outline below the next steps to operate predictive control in a true closed-loop configuration. Implementing integral action on the predicted POL wavefront will provide a more stable performance while also limiting the number of tunable parameters, further simplifying the use of the algorithm. 

In a closed-loop configuration the error measured by the PyWFS, $e(t)$, is the difference between the input disturbance $d(t)$ and the corrected phase given by the control signal at the last time step, $V(t-1)$, as shown in Eq.~\ref{eq:error}.

\begin{equation}
e(t)=d(t)-V(t-1)
\label{eq:error}
\end{equation}

Assuming the PyWFS measures the true error signal perfectly we can write $e(t)$ as a function of the slopes as in Eq.~\ref{eq:error_slopes}. 

\begin{equation}
e(t)=CM \times S(t)
\label{eq:error_slopes}
\end{equation}
The predictive filter is predicting the full input disturbance, $d(t)$, that will occur when the commands are sent. We can plug Eq.~\ref{eq:error} into Eq.~\ref{eq:leaky_int} to get the closed-loop configuration of predictive control with integral action, Eq.~\ref{eq:pred_closed}. 
\begin{equation}
    V(t)=(k+g) V^(t-1)-gV(t)_{pred} +g (CM \times-S_{ref})
\label{eq:pred_closed}
\end{equation}

\section{Conclusions}
\label{sec:conclusions}

We present the latest results from predictive control on Keck II AO system in this paper. We show that predictive control:
\begin{enumerate}
    \item improves the on-sky SR of NIRC2,
    \item provides a repeatable contrast gain of a factor of 1.5-3 on-sky at small separations (3-7 $\lambda/D$) with the vortex coronagraph,
    \item provides a more reliable correction under different seeing conditions, 
    \item and the improved performance is not driven by seeing.
\end{enumerate}
 Close inspection of the RMS wavefront error during the exposure of the NIRC2 images used for contrast analysis shows a reduction in the standard deviations of the distribution of wavefront errors for predictive control, revealing that prediction provides a more consistent correction over these time periods. Studying the noise and contrast curves from daytime and night time data respectively, we show that predictive control provides a repeatable improvement in contrast over the leaky integrator implemented for Keck II's PyWFS. 

We also present: 
\begin{enumerate*}
\item a working implementation of RLMMSE during daytime tests,
\item a more optimized EOF predictor using a lag of 2 frames and $\alpha$ equal to one,
\item three different nights showing a reduction in RMS wavefront error with predictive control, and
\item a clear path to future controller upgrades to improve upon our predictive control implementation.
\end{enumerate*}

\appendix    
\section{Contrast Gains}
\begin{figure}
    \centering
  \includegraphics[width=0.95\textwidth,trim={1cm 2.5cm 0cm 2.5cm},clip]{ 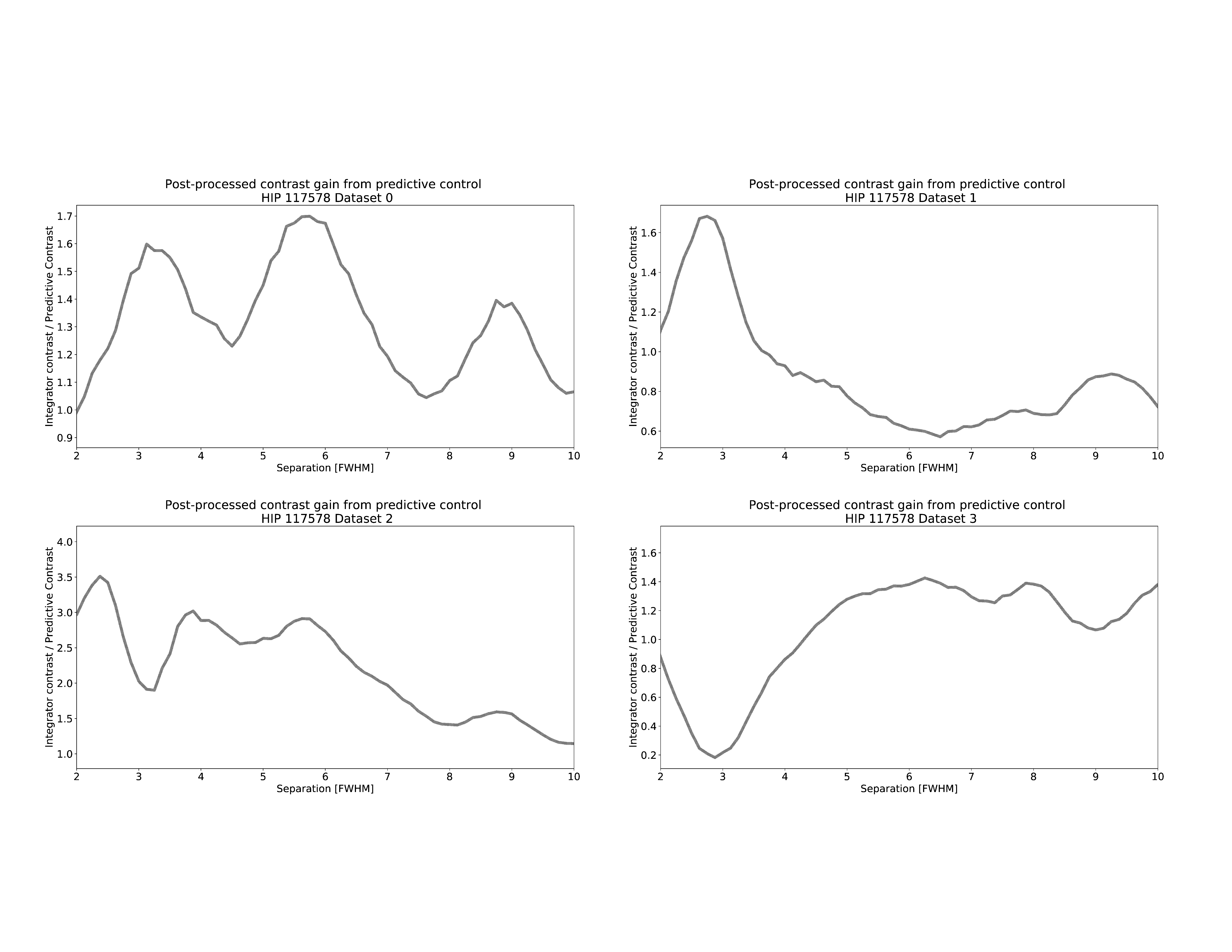}
 \caption{Contrast gains for HIP 117578 calculated by dividing the ADI contrast from the integrator by the ADI predictive contrast. Values greater than 1 demonstrate better performance with the predictive controller. The datasets correspond to the datasets in Tab.~\ref{tab:July28_parameters}}.
    \label{fig:contrast_gains_july28}  
\end{figure}

For further reference, we show the contrast gains achieved for the four different datasets in Fig.~\ref{fig:contrast_gains_july28}. These curves were calculated by dividing the contrast curves shown in Fig.~\ref{fig:contrast_july28}. 

\label{app:gains}

\acknowledgments 
The predictive wavefront control demonstration is funded by the Heising-Simons Foundation. The W. M. Keck Observatory is operated as a scientific partnership among the California Institute of Technology, the University of California, and the National Aeronautics and Space Administration. The Observatory was made possible by the generous financial support of the W. M. Keck Foundation. The near-infrared pyramid wavefront sensor (PyWFS) development was supported by the National Science Foundation under Grant No. AST-1611623. The PyWFS camera was provided by Don Hall as part of his National Science Foundation funding under Grant No. AST 1106391.  This work benefited from the NASA Nexus for Exoplanet System Science (NExSS) research coordination network sponsored by the NASA Science Mission Directorate. The authors wish to recognize and acknowledge the very significant cultural role and reverence that the summit of Maunakea has always had within the indigenous Hawaiian community. We are most fortunate to have the opportunity to conduct observations from this mountain.

This work builds on our results presented at the SPIE conference by van Kooten et al. (2021)~\cite{vanKooten_2021}.
\bibliography{report} 
\bibliographystyle{spiebib} 

\section{Tables}
\begin{table}[ht]
\centering
\begin{tabularx}{\textwidth}{|| c | X| X| X| X| X| c| c| c| c| c||} 
 \hline 
  &\multicolumn{5}{|c|}{Measured Performance} &\multicolumn{5}{|c|}{Controller Settings} \\
  \hline
 Dataset & \small{filter} $\frac{{Med_{int}}}{Med_{pred}}$  &\small{filter} $\frac{Std_{int}}{Std_{pred}}$ &\small{full} $  \frac{{Med_{int}}}{Med_{pred}}$ & \small{full} $\frac{Std_{int}}{Std_{pred}}$ &  $\frac{C_{int}}{C_{pred}}$ & $g$ & $m$ & HO gain & CoF & Order \editsA{($n$)} \\
 \hline \hline
 0 & 1.1 & 0.9 & 1.0 & 1.0 &1.5 & 0.15 &0.56 & 0 & 3 & 10 \\ 
 \hline
 1 & 1.1 & 1.0 & 1.0 & 1.0 &1.6 & 0.3 & 0.5 & 0.35 & 3 & 10  \\
 \hline
 2 & 1.1  &1.7 & 1.0 & 1.5 &2.0 & 0.3 & 0.5 & 0 & 5 & 10  \\
 \hline
 3 &  1.2 &  0.8 & 1.0 & 0.9 & 0.2 & 0.3 & 0.57 & 0 & 5 & 20\\
 \hline 
\end{tabularx}

\caption{The ratios of the median (med) and standard deviation (std) of $CL_{filt}$ and $CL_{full}$ (indicated by `filter' and `full', respectively) for each data set along with the ratio of contrast (C; contrast gain) at a separation of 3 $\lambda/D$.  For each measured performance metric a value greater than 1 indicates the predictor is performing better than the integrator. The controller settings for each of the data sets are shown and the number of modes controlled for all data sets was 300 modes with the LO gain being set to 1. For the NIRC2 images, the exposure was set to 0.3 seconds and the co-adds to 90 for all images. \editsA{The temporal order (Order) and cut-off-frequency (CoF) are also given for each data set.} These values correspond to the histograms shown in Fig.~\ref{fig:hist_July28}~\editsA{showing data taken the night of July 27, 2021}.}
\label{tab:July28_parameters}
\end{table}

\listoffigures
\end{document}